\documentclass[journal]{IEEEtran}
\usepackage{amsmath,amsfonts}
\usepackage{algorithmic}
\usepackage{algorithm}
\usepackage{array}
\usepackage[caption=false,font=normalsize,labelfont=sf,textfont=sf]{subfig}
\usepackage{textcomp}
\usepackage{stfloats}
\usepackage{url}
\usepackage{verbatim}
\usepackage{graphicx}
\usepackage{cite}
\usepackage{kotex}
\usepackage{subfig}
\usepackage{authblk}
\usepackage{multirow}
\usepackage{underscore}
\usepackage[english]{babel}

\graphicspath{ {./images/}} 

\begin{document}
\title{Searching for Effective Preprocessing Method and CNN-based Architecture with Efficient Channel Attention on Speech Emotion Recognition}

\author{Byunggun Kim, Younghun Kwon
\thanks{This work was supported by the Basic Science Research Program through the National Research Foundation of Korea (NRF), funded by the Ministry of Education, Science, and Technology (NRF2022R1F1A1064459). \\
\indent Byunggun Kim is with the Department of Applied Artificial Intelligence, Hanyang University(ERICA), Ansan, Kyunggi-Do, 425-791, Republic of Korea \\
\indent Younghun Kwon is with the Department of Applied Artificial Intelligence and Department of Applied Physics, Hanyang University (ERICA), Ansan, Kyunggi-Do, 425-791, Republic of Korea.
}}

\maketitle

\begin{abstract}
Speech emotion recognition (SER) classifies human emotions in speech with a computer model.
Recently, performance in SER has steadily increased as deep learning techniques have adapted.
However, unlike many domains that use speech data, data for training in the SER model is insufficient.
This causes overfitting of training of the neural network, resulting in performance degradation.
In fact, successful emotion recognition requires an effective preprocessing method and a model structure that efficiently uses the number of weight parameters.
In this study, we propose using eight dataset versions with different frequency-time resolutions to search for an effective emotional speech preprocessing method.
We propose a 6-layer convolutional neural network (CNN) model with efficient channel attention (ECA) to pursue an efficient model structure.
In particular, the well-positioned ECA blocks can improve channel feature representation with only a few parameters.
With the interactive emotional dyadic motion capture (IEMOCAP) dataset, increasing the frequency resolution in preprocessing emotional speech can improve emotion recognition performance.
Also, ECA after the deep convolution layer can effectively increase channel feature representation.
Consequently, the best result (79.37UA 79.68WA) can be obtained, exceeding the performance of previous SER models.
Furthermore, to compensate for the lack of emotional speech data, we experiment with multiple preprocessing data methods that augment trainable data preprocessed with all different settings from one sample.
In the experiment, we can achieve the highest result (80.28UA 80.46WA).

\end{abstract}


\section{Introduction}
\IEEEPARstart{S}{peech} emotion recognition (SER) is the technique in which a computer can recognize the inherent emotional features of a human's speech signal \cite{schuller2018speech}.
In particular, interest in the human-computer interaction (HCI) systems has arisen \cite{brave2007emotion}, and SER has noticed that some applications, such as psychotherapy \cite{yang2012detecting} and consultation calls \cite{bojanic2020call}, require emotional labor.
However, it is hard to understand how human speech emotions can be represented in terms of the exact values using common standards.
This is because each person's method of recognizing emotions differs depending on their personality and culture \cite{russell1991culture}.
The ambiguity of human emotions is one of the main challenges in developing an accurate SER model \cite{jahangir2021deep}.

Therefore, many studies have attempted to use the deep learning-based models \cite{han2014speech, fayek2017evaluating, mirsamadi2017automatic, xie2019speech, cao2021hierarchical}, that can be trained directly from the emotional speech data.
Among them, in recent, the convolutional neural network (CNN) based models trained with speech spectral features are proposed \cite{lim2016speech, satt2017efficient, ma2018emotion, chen20183, meng2019speech, peng2020speech, aftab2022light}. 
A CNN-based model has two advantages.
First, owing to the convolutional layer's weight sharing, a relatively small number of trainable parameters can be used.
Second, a deep CNN layer model makes it possible to learn global context features using filters of only a small size \cite{simonyan2014very, islam2020much, han2020contextnet}.
For the SER, it is essential to learn the linguistic features and the paralinguistic features of speech \cite{wani2021comprehensive}. 
Therefore, the CNN model, which can learn the context of emotional speech utterances, performs better than the other structures.

However, to learn the overall context of the input data with a CNN-based model, a sufficient number of deep layers must be stacked, or a larger filter kernel must be used.
Therefore, attention modules have been proposed to cover the CNN layer's weakness for the SER \cite{xu2020improve, li2021spatiotemporal, xu2021speech, liu2022atda, guo2023dstcnet, chen2023learning}.
Xu et al. \cite{xu2021speech} proposed a multiscale area attention, which applies the transformer-type attention mechanism \cite{vaswani2017attention} to the CNN-based model.
This significantly improves the recognition performance by dividing the time-frequency spatial information into granular perspectives.
Guo et al. \cite{guo2023dstcnet} proposed spectral temporal channel attention, which is a modified version of bottleneck attention module (BAM) \cite{hu2018squeeze, park2018bam, park2020simple}.
Therefore, it used not only focus on spatial features but also attention to channel features.
In addition, it has an independent attention learning structure in all the axes of the input features.
However, channel attention requires more learning parameters than spatial attention because of the two multi-layer perceptron (MLP) layers.

More trainable parameters are required when examining the attention structure and considering the more diverse aggregated input features \cite{hu2021model}.
However, an increase in trainable parameters causes overfitting problems when trainable samples are leaked, such as in SER \cite{el2011survey}.
Therefore, in this study, we search for an efficient attention structure that can improve emotional feature expression with only a few learning parameters while maintaining a deep CNN-based model.
Through experiments, we observe that CNN’s channel features are essential for emotion classification performance.
Therefore, the efficient channel attention (ECA) \cite{wang2020eca} that can learn how to focus on the important channel features is applied to the SER problem for the first time.
With the interactive emotional dyadic motion capture (IEMOCAP) corpus \cite{busso2008iemocap}, we experiment to look for methods to use the ECA with a CNN-based model that is more suitable.
The ECA can find important channel features by focusing on the relationship between neighboring channels for the emotion classification.
To achieve this, the ECA uses a 1-D convolution layer.
Therefore, it is highly efficient because it requires only a few trainable parameters equal to the kernel size \cite{kiranyaz20211d}.

\begin{figure*}[!t]
\centering
\includegraphics[width=7in]{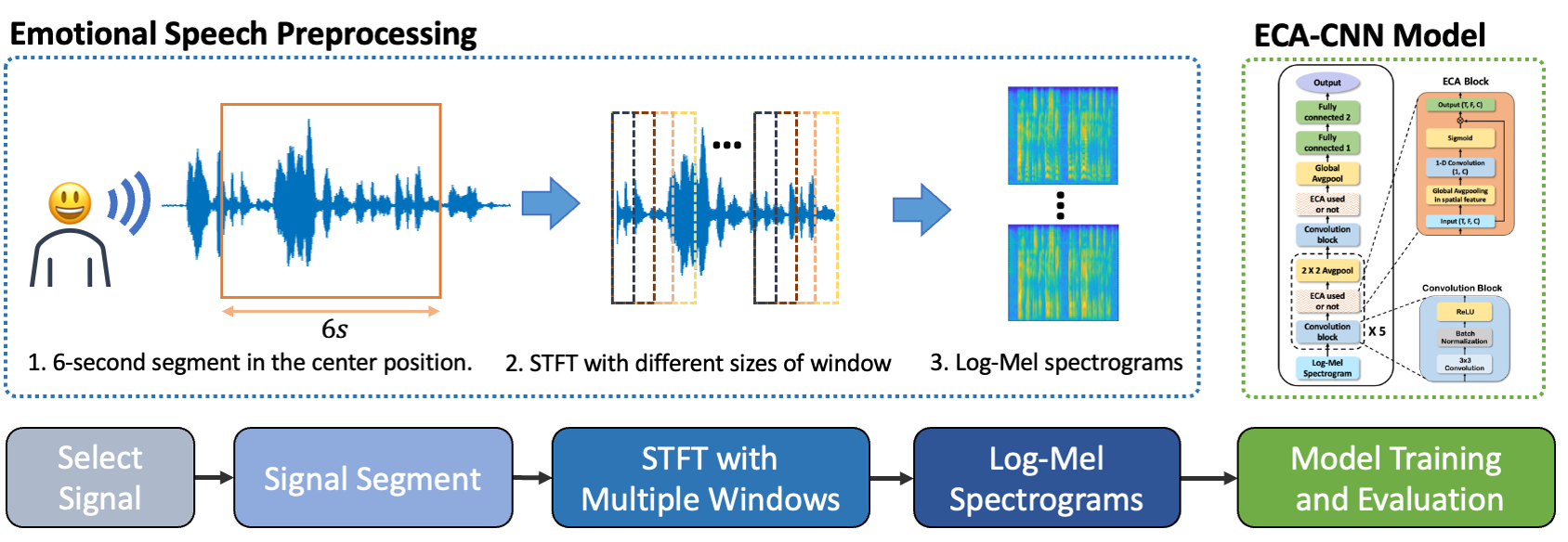}
\caption{The overall pipeline of speech emotion recognition with CNN-based efficient channel attention architectures.}
\label{overall_pipeline}
\end{figure*}

We also search for a more appropriate preprocessing method to better represent the emotional features.
Previous studies have preprocessed emotional speech signals using different methods; therefore, it is difficult to compare the results.
Therefore, we prepare the eight different types of preprocessed datasets.
Specifically, we choose the log-Mel spectrogram preprocessing method first.
We preprocess the speech signals with different and overlapping window sizes using short-term Fourier transformation (STFT).
We also evaluate the emotion classification performances of various CNN-based models.
As a result, we can observe that a preprocessing method with a large window size can accurately represent the emotional features.
Fig. \ref{overall_pipeline} shows the overall pipeline used in the experiments.

In summary, our contributions to this paper are below:
\begin{enumerate}{}{}
\item{Several experiments have shown that a CNN's channel complexity significantly affects the SER.
Therefore, the ECA was applied to the SER domain for the first time.
The ECA efficiently improved the emotion classification results with only a few training parameters by extracting the relationship between the features of the neighboring channel.
It was shown, in particular, that using ECA is effective in the deep layer of the CNN model, which has many channels.
}
\item{We conducted experiments using eight datasets preprocessed under various settings to determine the most effective preprocessing method for accurately expressing emotional features.
The precision of our results, a testament to the reliability of our experiment's outcomes, was evident.
Training with a log-Mel spectrogram, representing a relatively high-frequency resolution, proved significantly more effective in emotion classification.}
\item{In the field of SER, the need to learn data poses a significant challenge.
To address this, we proposed STFT data augmentation, which uses various preprocessed settings in STFT to supplement the expression of emotional features.
Our proposed STFT data augmentation had a profound impact, resulting in a substantial enhancement in emotion classification performance, demonstrating the efficacy of this approach.
When using a CNN-based model with ECA applied, we achieved the highest performance to date (80.28UA 80.46WA 80.37ACC), inspiring confidence in this method's potential.}
\end{enumerate}

This paper's overall composition is as follows: \\
Section \ref{related_works} describes the several CNN-based models with attention modules proposed for SER. 
Section \ref{preprocessing method} and Section \ref{model_architecture} introduce our proposed method. 
Section \ref{experiemnts_and_results} presents the details of the experimental settings and evaluation results when the ECA is applied to our CNN-based model. 
Finally, Section \ref{conclusion} concludes the paper.

\section{Related works}
\label{related_works}
Many different attention methods have been proposed.
In this section, we present an overview of the development flow of CNN-based models using several attention methods.
We divide contents whether the recurrent neural network (RNN) is used or not in the CNN-based model.

\subsection{CNN-RNN Models with Attention Mechanism}
The CNN can effectively learn local spatial features from the data. 
It is also possible to learn the global spatial features, such as the context of the data when stacking multiple CNNs. 
However, if the model stacks more layers, its complexity increases. 
In the SER problem, increasing the model complexity is critical. 
Therefore, to train the emotional context from the speech signals while sustaining the model complexity, most of studies have proposed a combination model a CNN and RNN \cite{lim2016speech, satt2017efficient, ma2018emotion}.
Although it is possible to learn the temporal features of speech using RNN, there are limitations to learning long sequences.
Therefore, to compensate for the limitations, many models that combine attention mechanisms with RNN have been proposed.

M. Chen et al. \cite{chen20183} proposed a CNN-LSTM-Attention model to aggregate the hidden states in each time step.
This enables effective learning even for long sequences. 
In addition, for more comprehensive context learning, ADRNN \cite{meng2019speech}, which uses residual connections and dilated convolution layers together, and ASRNN \cite{peng2020speech}, which compensates for the shortcomings of RNN through a sliding RNN method, have been proposed.

However, models using CNN-RNN-Attention layers forcefully learn the spatio-temporal features together. 
However, models consisting of CNN and RNN models in parallel are proposed to independently learn spatio-temporal features \cite{jiang2019parallelized}.
Zhao et al. \cite{zhao2019exploring} proposed a structure that separates LSTM and CNN in parallel and applies independent attention. 
Furthermore, Z. Chen et al. \cite{chen2023learning} proposed AMSnet, which is a parallel model that effectively synthesizes features through a connection attention mechanism.

\subsection{CNN-based Models with Attention Mechanism}

Recently, there has been a trend to use only CNN and attention layers without an RNN. 
Because RNN requires many more computational and training parameters than others, they focus on developing attention mechanism methods that help learn the context of the spatial features of speech spectrogram.
Xu et al. \cite{xu2020improve} demonstrated effective emotional feature learning only through self-attention after the CNN layer for the spatial context learning. 

Another attention mechanism method was proposed to enhance the feature learning of the CNN layers. 
Li et al. \cite{li2021spatiotemporal} demonstrated the importance of frequency features that use frequential attention, in addition to spatial attention. 
Xu et al. \cite{liu2022atda} proposed the ATDA, which applied independent self-attention to all feature axes (temporal, frequential, and channel) to compensate for the weakness of temporal feature learning owing to the lack of an RNN. 
Guo et al. \cite{guo2023dstcnet} proposed STC, which is a more efficient attention method for all feature axes of the CNN structure.

We further explore this trend and attempt to find an attention structure that is efficient and effective in learning emotional features while using a deep-layer CNN structure.
In particular, for efficient attention, we focused on learning the channel features that contained the context information of the input data in the CNN layer.
Therefore, we applied the ECA module to the SER problem and obtained improved emotion classification performance than before.

\section{preprocessing method}
\label{preprocessing method}

In this section, we explain the preprocessing method used to extract crucial emotional speech signal features.
Speech preprocessing suitable for a specific purpose is necessary to effectively learn a neural network model that may provide better performance.
It is not yet known which speech preprocessing method is the best for emotion recognition in speech.

Therefore, we need to determine which speech preprocessing method is the best way to recognize the emotion of speech.
For this purpose, we selected the log-Mel Spectrogram, which is frequently used for speech recognition.
Then, we check the suitable windowing and overlap times in the log-Mel Spectrogram to obtain the best result for the emotional recognition of speech.

\subsection{STFT}
The STFT is a method used to obtain the feature in frequency features by dividing a signal into short time periods.
Even though STFT is used in speech processing, the setting in STFT that is the most suitable for the emotional recognition of speech has yet to be discovered.
Therefore, we want to determine the value of the best setting when a neural network based on a CNN is applied to speech emotion recognition.
The STFT can be described as follows:

\begin{equation}
\label{stft}
X_t(f)=\sum_{\tau=-\infty}^{\infty} x(\tau)w(\tau - ts)e^{2\pi i f \tau}
\end{equation}

\begin{equation}
\label{stride}
s = l - o
\end{equation}

The output $X_t(f)$ is obtained by applying a windowing function $(w)$ to a signal $x(\tau)$ in an interval of the windowing function.
The output $X_t (f)$ then becomes a feature of the frequency.
The windowing function moves according to stride $(s)$, which is determined by the windowing length $(l)$ and overlapping length $(o)$.

\begin{figure}[!t]
\subfloat[\label{spectro_sad}]{
    \includegraphics[width=0.23\textwidth]{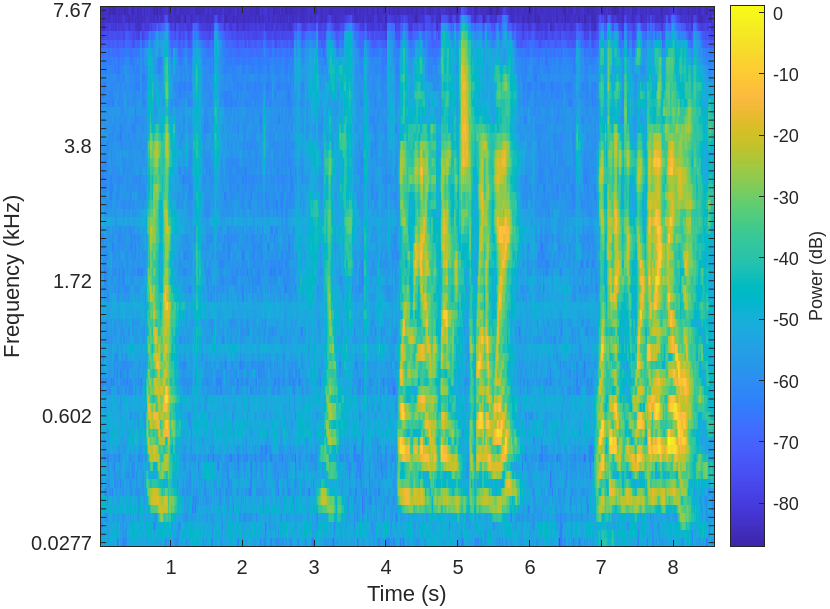}
}
\hfill
\subfloat[\label{spectro_hap}]{
    \includegraphics[width=0.23\textwidth]{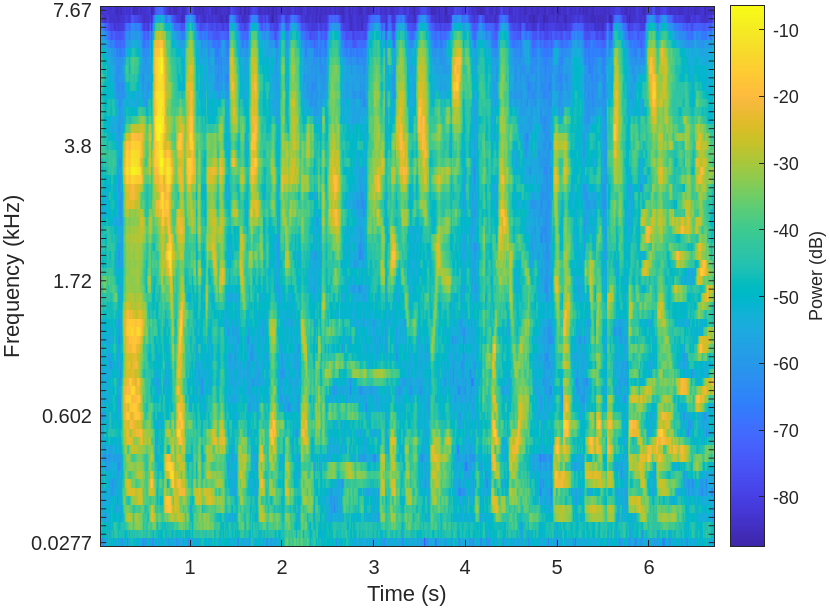}
}
\hfill
\subfloat[\label{spectro_ang}]{
    \includegraphics[width=0.23\textwidth]{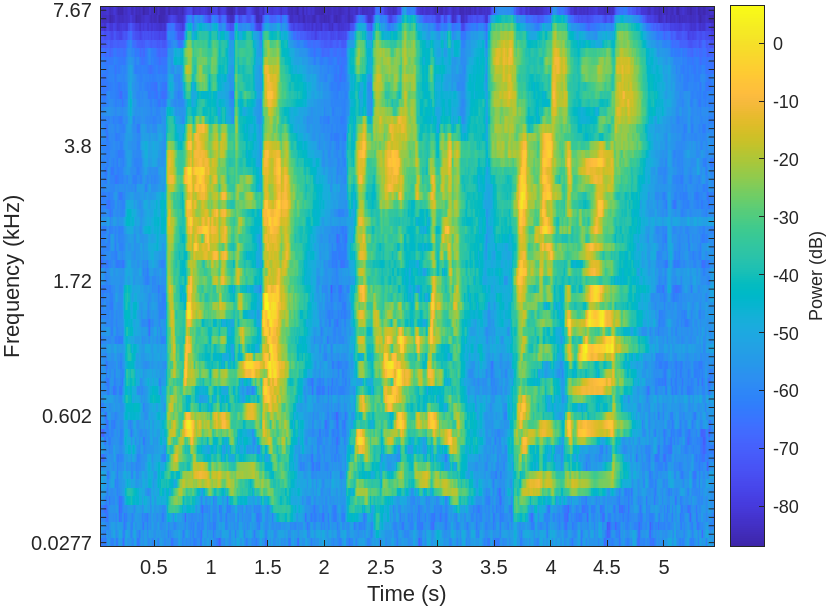}
}
\hfill
\subfloat[\label{spectro_neu}]{
    \includegraphics[width=0.23\textwidth]{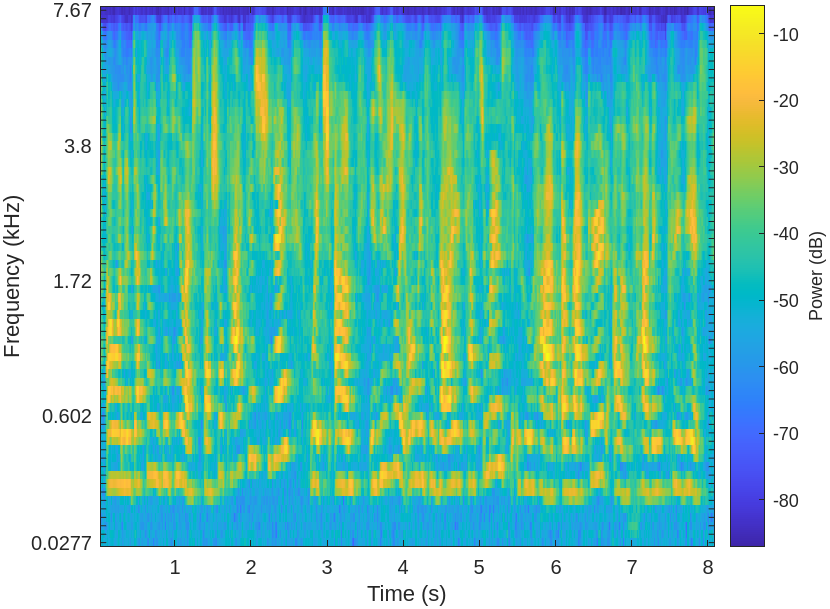}
}
\caption{The log-Mel spectrogram images of each category of emotions preprocessed by the setting of version 8. (a) Sadness (Ses05M impro02 M013). (b) Happiness (Ses05M impro03 M018). (c) Angry (Ses03M impro05a M012) (d) Neutral (Ses02F impro08 F005)}
\label{emotion_spectrogram}
\end{figure}

Note that the output $X_t (f)$ has a resolution limit.
If the windowing length is longer, the frequency resolution increases; however, the resolution in time decreases. 
If the windowing length is shorter, the frequency resolution decreases, however, the time resolution increases.
Therefore, we need to determine which features are more important in terms of time or frequency.
For this purpose, we performed our experiment by using eight different settings during preprocessing.

\subsection{Speech Emotion Discrimination with Log-Mel Spectrogram}

Because the amount of data for emotion recognition in speech is limited, it is advantageous for the size of the features to become small.
The log-Mel spectrogram is effective for emotion recognition in speech because it can reduce the size of the features expressed in frequency.
In addition, the log-Mel spectrogram displays speech characteristics in 2D images.

Fig. \ref{emotion_spectrogram} shows an example of a log-Mel spectrogram for each emotion class.
Fig. \ref{emotion_spectrogram} (a) and Fig. \ref{emotion_spectrogram} (c) show the characteristics of speech in sadness and angry emotions.
In the case of sadness, the utterance time was short, and the speech was distributed in the low-frequency regions. 
Meanwhile, some parts of the high-frequency region were observed in the speech of angry.

However, recognizing the differences between different emotions is challenging.
In addition, some images are ambiguous when characterizing the corresponding emotions.
Therefore, a deep neural network model based on a CNN must be introduced to recognize the emotions of speech. 
In this study, we suggest the structure of a CNN model based on efficient channel attention, which is effective and efficient in emotion recognition of speech.

\section{Model architecture}
\label{model_architecture}
Because there is little training data for speech-emotional recognition, we need to use as few parameters as possible to improve the successful recognition of emotions in speech.
Therefore, we consider a neural network model for images based on a CNN. 
In addition, to effectively learn speech emotions in context, we set up a model to elevate the learning ability in the channel. 
Therefore, we apply efficient channel attention (ECA) to our neural network model-based CNN.

\subsection{Convolution Layer}

\begin{table}
\begin{center}
\caption{parameter settings of each layer in baseline model}
\label{tab1}
\begin{tabular}{ c  c }
\hline
Layer & parameter settings \\
\hline
Conv1 & kernel size = (3,3), stride=1, number of filters = $16 * n$ \\
\hline
Conv2 & kernel size = (3,3), stride=1, number of filters = $32 * n$ \\
\hline
Conv3 & kernel size = (3,3), stride=1, number of filters = $48 * n$ \\
\hline
Conv4 & kernel size = (3,3), stride=1, number of filters = $64 * n$ \\
\hline
Conv5 & kernel size = (3,3), stride=1, number of filters = $80 * n$ \\
\hline
Conv6 & kernel size = (3,3), stride=1, number of filters = $96 * n$ \\
\hline
Linear1 & weight parameter shape = $(96*n, 96*n)$ \\
\hline
Linear2 & weight parameter shape = $(4, 96*n)$ \\
\hline
Avgpool & kernel size = (2,2), stride=2 \\
\hline
\end{tabular}
\end{center}
\end{table}

The Convolution layer learns the input features included in the local region by using various filters.
The 2D Convolution layer learns the spatial information of the 2D images.
When a 2D convolution layer is applied to a spectrogram for recognition for recognizing speech emotions, it can learn the relationship between time and frequency.

\begin{equation}
\label{convolution}
(X^l * W_k^l)_{ij} = \sum_t \sum_f \sum_c X^l(t, f, c) \cdot W_k^l(i-t,j-f,c)
\end{equation} 

\begin{equation}
\label{2d_conv}
\mathrm{2DConv}(X^{l+1})= [(X^l*W_1^l), ..., (X^l*W_{c'}^l)]
\end{equation}

Equations (\ref{convolution}) and (\ref{2d_conv}) show the method for calculating the convolution when a spectrogram is used as the input.
The input $X^l \in \mathbb{R}^{T \times F \times C}$ is convoluted with weight filter $W^l=[W_1^l, ... , W_{C'}^l ] \in \mathbb{R}^{K \times K \times C \times C'}$ to transform the $C'$ size of channel dimension.
In this calculation, the weight filters are used with kernel size $(K \times K)$ of the local spatial information.

We must set an appropriate number of the weight filters and kernel sizes to better train with this convolution layer.
Therefore, in this study, we explored the trainable parametric efficiency in a CNN-based model for speech emotion recognition.
We mainly focus on some weight filters called “channel size”. 
The channel size in the convolution layer is a hyperparameter related to the information capacity of the neural network. 
If we use many weight filters, spatial information can be sensitively extracted from the input data. 
However, this leads to a more robust overfitting of the training dataset.
We design a 6-layer CNN architecture to determine an adequate channel size for speech emotion recognition.
This is explained in detail in the following section.

\subsection{CNN-Based Architecture}

\begin{figure}[!t]
\centering
\includegraphics[width=2.5in]{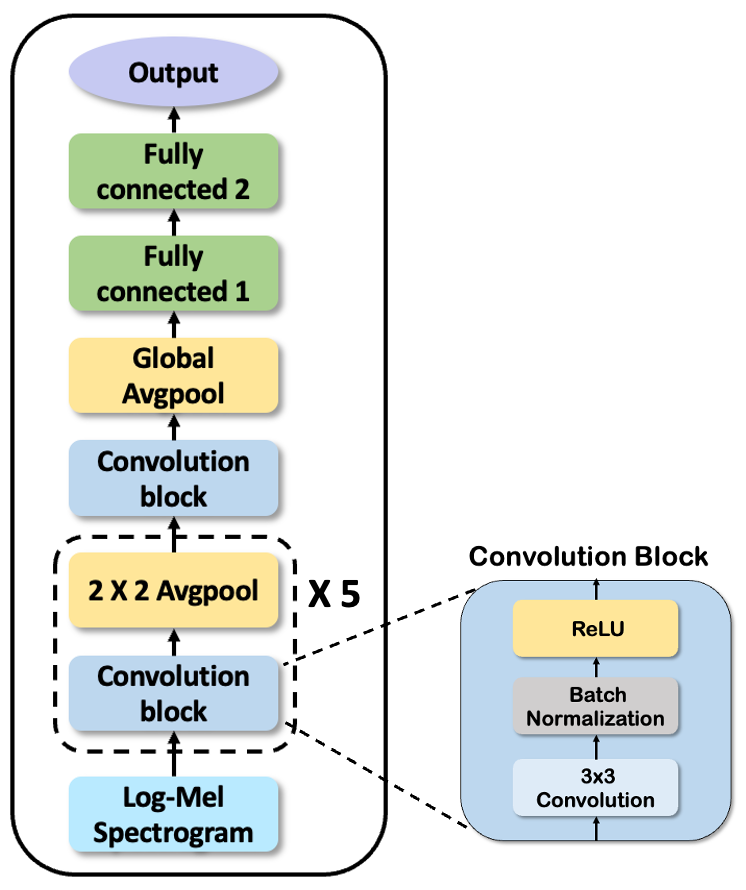}
\caption{CNN-based model architecture. It consists of six convolution blocks and two fully connected layers.}
\label{baseline_model}
\end{figure}

We proposed a deep CNN-based model as a baseline for developing an effective SER model.
The structure of the CNN-based model was based on a previously proposed model for SER. 
Using this model, we focused on CNN’s channel features of the CNN, which effectively trained the speech emotion features. 
Fig. \ref{baseline_model} shows the overall architecture of the CNN-based model. 
It is composed of six Convolutional blocks and a pooling layer. 
And, two fully connected layers are used for emotion classification.

The detailed model structure is as follows.
First, we chose a convolution block, which is commonly used in image classification models.
The convolution block consists of three layers: convolution layers with $(3 \times 3)$ kernel size, batch normalization, and ReLU activation.
To find the adequate channel size in each convolution layer, we selected the initial channel size in multiples of $16$. 
We multiplied by a scaling factor to scale up the channel size.
The details of the parameter settings are listed in Table \ref{tab1}.

Next, we used the average pooling layer after all the convolution blocks.
It can efficiently subsample the hidden features without vanishing the features.
Exceptionally, in the sixth pooling layer, we used global average pooling to connect with the next layer. 
Finally, two fully connected layers were used in the classification layer. 
The weight of each fully connected layer was set to be the same as the output channel size of the last convolution layer.

\subsection{The Design of ECA Module}

\begin{figure}[!t]
\centering  
\includegraphics[width=3in]{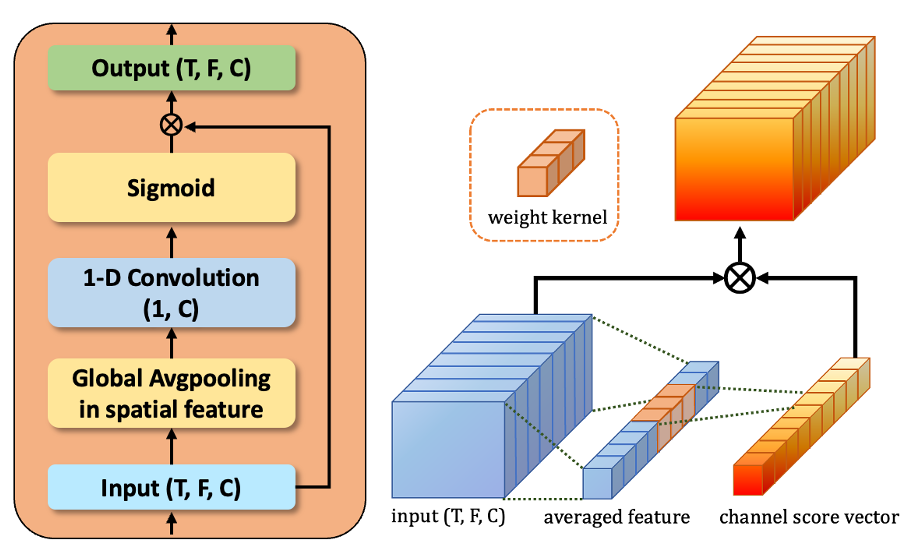}
\caption{The design of efficient channel attention module in this study.}
\label{eca_block}
\end{figure}

The convolution layer can extract the local spatial features using the number of trainable filters from the input data.
However, when more filters were used in the model, the representation capacity of the filters weakened.
To overcome this situation, we adopt the ECA \cite{wang2020eca} in the CNN-based model, which can effectively improve the representation of the filters.
The ECA is a channel attention architecture proposed by Wang et al.. Learning the relationship between different filters and focusing on the important ones is possible with fewer trainable parameters.
We experimented with the application of an ECA suitable for a CNN-based SER model.
As a result, increasing the filter’s representation in the convolution layer helps to extract emotional features.

The ECA is a type of layer that applies a self-attention mechanism \cite{vaswani2017attention}. 
The self-attention mechanism comprises three components: query ($Q$), key ($K$), and value ($V$).
These components are used to attend to the crucial information from input features.
Each component's role is as follows.
A query is a "question" about what is essential in the input features.
The key is "hint," which uses how similar the query is. 
This helps to find the most appropriate information in the input features.
The value is the "real answer" that exists in pairs with keys and is used as the output features of self-attention.

\begin{equation}
\label{attention}
\mathrm{Attention} (Q, K, V) = \mathrm{Softmax}( {{QK} \over {\sqrt{n}}} )V
\end{equation}

The most common self-attention mechanism in (\ref{attention}) is the inner-product attention. 
In the first step, each query is multiplied with the multiple keys to obtain the relevant score matrix.
In the second step, the score matrix is represented by a probability distribution within each query using the Softmax function.
In the final step, the output is represented by a linear combination of all the values with a probability distribution.
In summary, the output is the weighted value obtained that is most closely related to the query indirectly through the keys.

\begin{equation}
\label{eca}
\mathrm{ECA} (Q, V) = \sigma (\mathrm{1DConv(Q)}) \otimes V
\end{equation}

\begin{equation}
\label{eca_query}
Q = \mathrm{GAP}(X)
\end{equation}

ECA is simplified by omitting the key from the existing self-attention structure.
Equation (\ref{eca}) shows the progress of the ECA.
The most significant difference from self-attention in (\ref{attention}) is that there is no key; however, the importance of the value is judged by the score learned from the query itself. 
This simplified attention mechanism is suitable for application in the SER.

The ECA structure is shown in Fig. \ref{eca_block}.
The ECA structure can be divided into three steps.
The first involves the preparation of a query. 
The query represents the channel features in each input.
We used global average pooling, which can contain temporal-frequency features in each channel without trainable parameters. Please refer (\ref{eca_query}).

The next step is to make the score that represents the importance of each channel feature.
We trained the channel feature’s relation with a 1-D convolution layer to obtain the score. 
Specifically, the number of neighboring channel queries that train the relation with the target channel query is determined by the kernel size of the 1-D convolution layer.

Moreover, with the sigmoid function, we score within 0 to 1.
Next, a key difference from the original ECA is the omission of batch normalization. 
This change is made because batch normalization tends to extract global features of speech data rather than individual emotional features.

The final step is the re-representation of the input features. 
An element-wise product is performed on the input features and the attention channel score.

\begin{figure}[!t]
\centering
\includegraphics[width=3in]{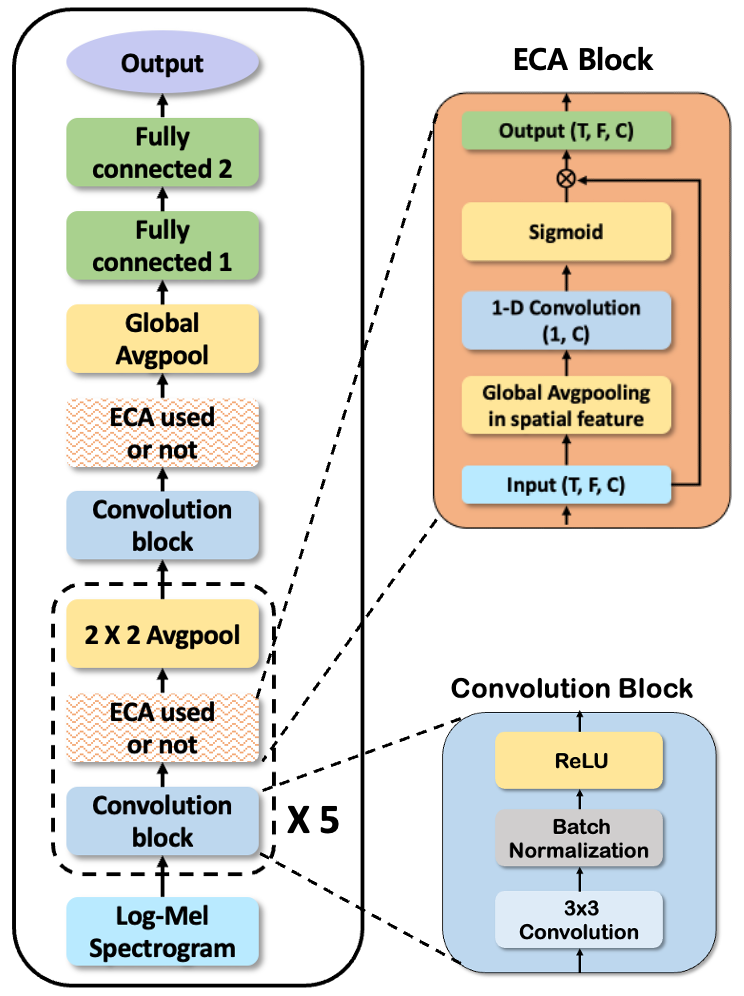}
\caption{The ECA block position in CNN-based model.}
\label{eca_cnn_model}
\end{figure}

ECA can efficiently improve the channel feature of the convolution layer, which is essential for classifying speech emotions.
Fig. \ref{eca_cnn_model} shows the ECA block used after the convolution block.
We searched for more appropriate ECA block settings for the CNN-based model.
To achieve this, we compared the different experiments using several kernel sizes and block positions in the model.
Consequently, unlike the original ECA method used in all layers after the convolution block, using some convolution blocks with many channel features can be effective for emotion recognition performance.
This shows that the ECA works well when the complexity of the filter increases.

In particular, the effectiveness of ECA can be increased when augmentation data are used for training.
Typically, sufficient data are required to use an attention-structured neural network effectively.
Therefore, we used eight different datasets as the augmentation data, as listed in Table \ref{tab2}. 
Consequently, the performance difference between the model combined with the ECA and the model without the ECA is visible.

\subsection{Weighted Focal Loss}
Most speech emotion datasets have an unbalanced distribution, depending on the emotion labels. 
Therefore, when we use cross entropy as a loss function, the model can be focused on a relatively large number of specific emotion labels (neutral and happiness). 
In addition, emotion classification is complex, depending on the label. 
For example, “happiness” and “angry” are often misclassified.
Therefore, we used weighted focal loss to deal with those specific situations.
The weighted focal loss can be expressed as follows.

\begin{equation}
\label{loss}
\mathrm{Weighted Focal} (p_i, w_i) = - \sum_{i=1}^{n_{class}} w_i (1-p_i)^{\gamma} \log (p_i)
\end{equation}

It has two properties.
First, the focal loss \cite{lin2017focal} achieves flexible learning rates in different emotional classes $(n_{class})$. 
The focal loss function is a slightly more generalized function of the cross entropy weighted by the predicted probability of each emotion class $(p_i)$. 
For emotion classes with low probability values, the loss increases.
This causes relatively more learning in backward updates. Conversely, emotion classes with high probability values resulted in relatively less learning.
The hyperparameter gamma ($\gamma$) controls how dramatically the loss value changes.
In our experiment, we set $\gamma=1$.

Second, to avoid imbalanced learning owing to the number of emotion classes, the loss function is multiplied by the learning rate weight $(w_i)$ based on the number of labels in the training dataset.
The learning rate weight was obtained as the reciprocal ratio of the number of data for each label.

\section{Experiment Setting and Results}
\label{experiemnts_and_results}

\subsection{Dataset (IEMOCAP)}
The IEMOCAP \cite{busso2008iemocap} is the most popular dataset used in the SER problem. 
In this dataset, ten individual actors (five men and five women) recorded their voices, facial movements, and overall behavior for five sessions to understand human emotions in various situations. 
The IEMOCAP's data are divided into an improvised set, which contains improvised acting, and a script, which acts through dialogue. 
Three or more annotators manually classified the emotional speeches, and the final decision was made through a majority vote.

The data samples that we used in the experiment are as follows.
We selected only the Improvised set, considering situations in which human emotions can appear naturally. 
Next, we choose five emotional classes corresponding to “angry”, “sadness", “happiness”, “neutral” and “excited”.
These classes are commonly used in various experiments. 
Moreover, to compensate for the data sample’s number of “happiness,” “excited” is considered as “happiness”.
The experiments were conducted using 2943 speech data samples containing four emotional classes (angry:289, sadness:608, happiness:947, neutral:1099).

\subsection{Data Preprocessing}

\begin{table}
\begin{center}
\caption{preprocessing setting in each version of datasets}
\label{tab2}
\begin{tabular}{ c  c  c  }
\hline
Dataset Version & Window & Overlap \\
\hline
1 & 15 ms & 5 ms \\
2 & 20 ms & 10 ms \\
3 & 25 ms & 15 ms \\
4 & 30 ms & 20 ms \\
5 & 35 ms & 25 ms \\
6 & 40 ms & 30 ms \\
7 & 45 ms & 35 ms \\
8 & 50 ms & 40 ms \\
\hline
\end{tabular}
\end{center}
\end{table}

\begin{table}
\begin{center}
\caption{STFT setting used in previous proposed SER models}
\label{tab3}
\begin{tabular}{ l  l  l  l  l }
\hline
Model Name & Method & Window & Overlap \\
\hline
ATDA \cite{liu2022atda} & MFCC & 48 ms & 24 ms \\
Area Attention \cite{xu2021speech} & log-Mel & 40 ms & 10 ms \\
MHA \cite{nediyanchath2020multi} & log-Mel & 46 ms & 23 ms \\
STC \cite{guo2023dstcnet} & spectrogram & 16 ms & 8 ms \\
HNSD \cite{cao2021hierarchical} & log-Mel & 25 ms & 10 ms \\
TIM \cite{ye2023temporal} & MFCC & 50 ms & 38.5 ms \\
AMSNet \cite{chen2023learning} & spectrogram & 50 ms & 25 ms \\
\hline
\end{tabular}
\end{center}
\end{table}

To efficiently represent the input data, we extracted the log-Mel spectrogram features from the speech signal. 
The data preprocessing steps are as follows. 

First, signal segmentation is performed to equalize the length of the input data. 
The IEMOCAP dataset speech samples had an average length of 4.5 s and varied from short ($\sim 0.5s$) to long ($\sim 30s$) speech. 
Therefore, the lengths of the speech samples were consistent at 6 s, which was slightly longer than the average. 
Specifically, if the speech data were shorter than 6 s, zero padding was performed at the beginning and end of the signal with the same length for the signal position in the center. 
If the speech was longer than 6 s, both the beginning and ends of the signal were cut to the same size to contain as long an utterance as possible.

Next, we prepared the different versions of the datasets to search for more effective preprocessing settings with different window sizes and overlaps in the STFT. 
Therefore, an interval was set based on previous studies. 
As listed in Table \ref{tab3}, most previous studies set the window size from 16 ms to 50 ms. 
Based on this, we chose eight different window sizes at 5ms intervals within a slightly wider range of 15 ms to 50 ms. 
The overlap size was adjusted to obtain the same size of input data.

In the final step, log-Mel filters were applied to effectively decrease the input data size
Specifically, we use 64-number Mel filters to increase frequency resolution. 
The detailed settings for each dataset version of the dataset are listed in Table \ref{tab2}.
All preprocessing steps were conducted using MATLAB.

\subsection{Experimental Setup and Evaluation}
We use 5-fold cross-validation for the entire dataset to ensure general SER performance. 
Samples from all the sets were randomly selected. 
To evaluate the model performances, we used the unweighted average accuracy (UA) and weighted average accuracy (WA). 
These two metrics are commonly used in the SER. 
Additionally, to analyze the balanced evaluation, we used the mean of UA and WA (ACC) values. 

The PyTorch framework \cite{paszke2019pytorch}, a deep learning framework, was implemented in all the experiments in this study. 
The specific hyperparameters of the models were as follows: 
All weight parameters were initialized with He initialization \cite{he2015delving}. 
For the model's optimization, we use the Adam optimizer \cite{kingma2014adam} with $10^{-4}$ initial learning rates and $10^{-6}$ decay rates. 
The batch size was set to 32, and the focal loss parameter $γ$ was set to $1$. 
Finally, the models were trained for 150 epochs.

\subsection{Searching the Proper Channel Size for CNN-Based Model Architecture}
\label{result1}

\begin{figure}[!t]
\centering  
\includegraphics[width=3.5in]{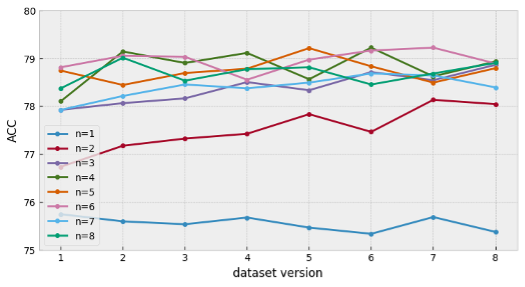}
\caption{The comparison of ACC results from models using different channel sizes and versions of the dataset.}
\label{cnn_channel_size_result}
\end{figure}

To extract emotional features from speech data more effectively, we first experimented with the number of channel sizes used in a CNN-based model.
The channel size is the most critical hyperparameter in the convolution layer. 
Also, the proper channel size is crucial to efficiently reduce the number of trainable weight parameters.
The optimal number of channels can improve the classification performance of the SER model.

For this experiment, we individually trained and evaluated the eight different CNN-based models with variant channel sizes. 
The detailed number of channel sizes is set with the parameter $n$, as listed in Table \ref{tab1}. 
Parameter $n$ was selected as an integer ranging from 1 to 8.
In addition, we used eight different datasets to analyze which version of the preprocessing methods best represents the emotional features. 
The preprocessing methods are listed in Table \ref{tab2}.

Fig. \ref{cnn_channel_size_result} shows the performance of the entire model for the different versions of datasets.
From the perspective of channel size change, the model performance changes drastically as the channel size increases in range from $1$ to $3$.
In particular, when comparing the performance of the $n=1$ (blue line) and $n=3$ (purple line) models, there was a $2 \% \sim 3 \%$ gap.

Next, the best result was obtained when we used the $n=4$ (khaki line) model with the dataset of version 6 (79.16 UA 79.27 WA 79.22 ACC).
In addition, the model with $n=4$ (khaki line) shows the best results for datasets of versions 2, 6, and 8. 
This shows that an appropriate channel size can lead to a decent performance in the emotion recognition possible.
However, no improvement was observed in the models using a larger channel size $(4\le n \le 8)$.
Therefore, a larger channel size can be inefficient for training. 

In experiments with different versions of datasets, except $n=1$, the best performance of each model can be observed in the higher versions of the datasets $(5 \sim 8)$.
This implies that a larger window size can effectively represent emotional features.

In summary, the channel size is an important factor in the performance of a CNN-based model. 
Therefore, in the next experiments, we explored how to improve the training channel features using the ECA.

\subsection{Using Original ECA Method in CNN-Based Models}

\begin{figure}[!t]
\centering  
\includegraphics[width=3.5in]{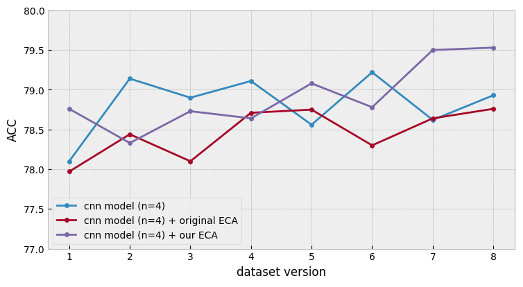}
\caption{The comparison of ACC results of the ECA block used or not in the CNN-based model with $n=4$. The proposed ECA block shows more effective results than the original ECA block in the version 7 and 8 datasets.}
\label{original_eca_result}
\end{figure}

To efficiently increase the channel feature representation, we use the ECA blocks in a CNN-based model. 
We first experimented with the original ECA block. 
For the experiments, we selected the CNN-based model ($n=4$) that achieved the best performance in a channel size search experiment in section \ref{result1}.
Moreover, the original ECA blocks in this model were applied.
The original ECA is positioned after each convolution layer.
Therefore six ECA blocks were added to the CNN-based model.
Next, the ECA’s kernel sizes $(k)$ were set based on the channel size of each layer.

Fig. \ref{original_eca_result} shows the performance of each dataset before and after applying ECA. 
Compared with the CNN-based model, the original ECA method (red line) showed an overall decrease in emotion recognition performance from $0.1\%$ to $1\%$.
This implies that the original ECA method is not suitable for a direct application. 
Therefore, we experimented to determine a more appropriate way to use the ECA block. 

The differences between the original ECA and our ECA methods are as follows:
First, we applied the ECA blocks after the 5th and 6th convolution layers, which were relatively deeper than the other layers. 
In addition, a larger kernel size $(k = 7)$ was used compared to the original kernel size $(k = 3)$. 
By applying our ECA, the number of trainable parameters increased by only 14.

Consequently, unlike the original ECA method, the proposed ECA method (purple line) obtained a $0.3\%$ higher performance than that without an ECA block.
In particular, the best performance (79.37 UA 79.68 WA 79.53 ACC) was obtained when we trained with dataset version 8. 
This result indicates that the ECA method can improve performance using SER properly. 
The channel features of deeper layers are particularly important for extracting the speech emotion features.

\subsection{Searching the Proper ECA Block Usage with Different Version of Datasets}

\begin{figure}[!t]
\centering  
\includegraphics[width=3.5in]{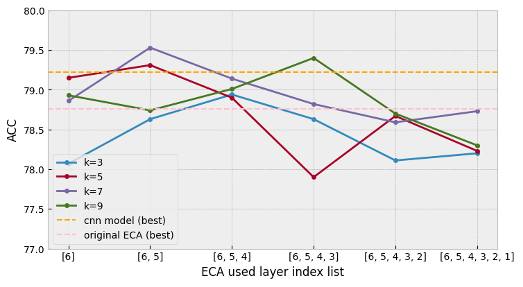}
\caption{The comparison of ACC results with the ECA blocks used in different layers of CNN-based model with different kernel sizes.}
\label{eca_search_result}
\end{figure}

To effectively use ECA blocks, we experimented with how the model performance changes when using the ECA blocks in different positions on a CNN-based model with different kernel sizes $(k)$.
A CNN-based model has a structure in which the channel size increases as the layer deepens; therefore, the complexity of the channel features in deep layers is relatively high.
Based on that situation, we want to determine where the ECA block can be more effective in helping to train the channel features.
Specifically, the experiment was conducted by sequentially adding the ECA blocks starting from the sixth convolution layer, which was the deepest layer in the model.

Subsequently, we changed the kernel size of the 1-d convolution layer of the ECA block.
The kernel size is the length of the local region in which the relationships between neighboring channel features are learned.
In our experiments, four different kernel sizes (3, 5, 7, and 9) were used to verify the change in performance according to the kernel size.
Moreover, we excluded the cases in which the kernel size was larger than nine because of poor performance.
However, to determine the best kernel size, we must train $(4+1)^6$ times to consider all case sizes according to the ECA block.
Therefore, we skipped some cases of the experiments owing to limitations in computational resources.
In all the experiments, we used the dataset of version 8.

Fig. \ref{eca_search_result} shows that the performance decreases when ECA blocks are used after most convolution layers, whereas the performance improves when ECA blocks are used for relatively deep layers (3 to 6 layers).
In particular, when ECA blocks are used on the fifth and sixth floors, the highest performance of the model without the ECA block (orange line, 79.16 UA 79.27 WA 79.22 ACC) was improved by approximately $0.3\%$ based on the ACC.
In addition, it shows high performance in many cases compared to the best result when the original ECA blocks are used (pink line, 78.71 UA 78.80 WA 78.76 ACC).
These results show that the ECA block positioned in the deep-layer channel features was effective.

Next, if the results are analyzed according to the kernel size, the kernel size that shows the best performance is $k=7$ (purple line). 
Compared with the other kernel sizes (3, 5, and 9), $k=7$ had an overall high accuracy range ($78.59 \sim 79.53$ ACC). 
However, $k=7$ did not always exhibit the best performance in any of the cases. 
For example, the four ECA blocks used in model $k=9$ (khaki line) performed better than the others (1, 3, and 5).

\begin{figure}[!t]
\centering  
\includegraphics[width=3.5in]{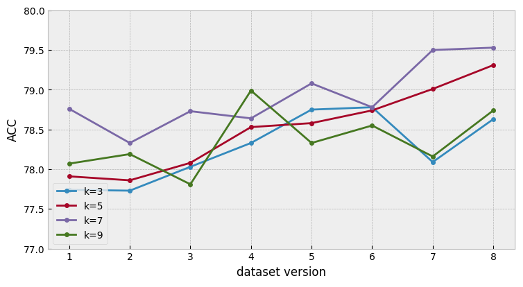}
\caption{The comparison of ACC results used different versions of datasets with the CNN-based model using the ECA blocks in layers 5 and 6.}
\label{n4_eca_result}
\end{figure}

In the next experiment, we examined the effect of the ECA block using eight different emotional speech preprocessing datasets.
For this purpose, a CNN-based model with the ECA blocks on the fifth and sixth convolution layers was used, which were the cases with the highest performance in the previous experiment.
In addition, four kernel sizes (3, 5, 7, and 9) were used.

As shown in Fig. \ref{n4_eca_result}, the model performance tended to increase from dataset versions 1 to 8
In particular, as in the previous experiment, the version 8 dataset showed better results than the other version datasets in most cases.
This indicates that a large-sized window in emotional speech preprocessing is effective.
However, for most dataset versions, the performance was lower than that of CNN-based models alone.
Eventually, it will be essential to train the model using fined emotional features to obtain an effect from the attention layer.

\subsection{Augmentation Method with Different Version of STFT Datasets}

To overcome the limitations of representing speech emotional features obtained using only one preprocessing method, multiple preprocessing data augmentation experiments were performed.
For this purpose, only the training dataset was added from the eight different versions of the datasets obtained by setting listed in Table \ref{tab1}.
Because each of the eight preprocessing methods has a different window size and overlap size, the model can train with richer emotional features.

\begin{figure}[!t]
\centering  
\includegraphics[width=3.5in]{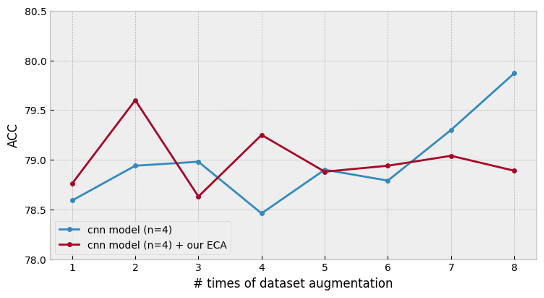}
\caption{The comparison of ACC results used the STFT data augmentations. The number of dataset are selected with dataset versions in ascending order.}
\label{ascending_augment}
\end{figure}

\begin{figure}[!t]
\centering  
\includegraphics[width=3.5in]{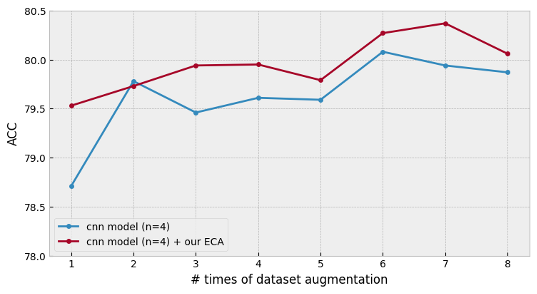}
\caption{The comparison of ACC results used the STFT data augmentations. The number of dataset are selected with dataset versions in descending order.}
\label{descending_augment}
\end{figure}

We conducted two different experiments depending on the dataset selection methods to determine out the effect of multiple preprocessing data augmentation on SER. 
In the first case, we selected version 1 as the test set and collected training data samples from version 2 to version 8 in ascending order.
Second, in contrast to the first, we selected dataset version 8 as the test set and collected the training data samples in descending order from version 7 to version 1.
The models used in the experiment are CNN-based models with ECA blocks and models without an ECA block.

Fig. \ref{ascending_augment} shows the augmentation experiments in ascending order.
In most cases, the results were higher than those in the cases where the augmentation method was not applied to either model.
Specifically, the best results in the CNN-based model were obtained using all the preprocessing datasets (80.10 UA 80.02 WA 80.06 ACC).
In the case of models with ECA blocks, the best results (79.69 UA 79.51 WA 79.60 ACC) were obtained when versions 1 and 2 were used.
These results show that the multiple preprocessing augmentation method can improve performance when a small amount of data is available.

Fig. \ref{descending_augment} shows the augmentation experiments in descending order.
An interesting result shown in Fig. \ref{descending_augment} is that the CNN model (79.66 UA 80.50 WA 80.08 ACC) and the ECA CNN (80.28 UA 80.46 WA 80.37 ACC) model exhibited the highest performance.
In particular, the CNN-based model with ECA blocks showed 0.8\% higher based on ACC than the ascending experimental result.
In addition, the CNN-based model with ECA blocks (red line) is usually 0.2\% to 0.82\% higher than that of the CNN model (blue line).
In other words, the multiple preprocessing augmentation method can significantly improve the learning of emotional features using ECA blocks.

From these two experiments, we can observe that the multiple preprocessing augmentation method can compensate for the training problem with a few speech-emotional data samples, which is one of the difficulties in SER.
In addition, the ECA blocks can work more effectively using the data augmentation method.
This method achieved the highest performance (80.28 UA 80.46 WA 80.37 ACC).

\subsection{Comparison with Other Attention Models}

\begin{table}
\begin{center}
\caption{performance comparison with different SER models}
\label{tab4}
\begin{tabular}{ l  l }
\hline
Model Name & Evaluation Results \\
\hline
STC \cite{guo2023dstcnet} & 59.1 UA 60.5 WA \\ 
AMSNet \cite{chen2023learning} & 70.5 UA 69.2 WA \\ 
MHA \cite{nediyanchath2020multi} & 70.1 UA 76.4 WA \\
HNSD \cite{cao2021hierarchical} & 72.5 UA 70.5 WA \\ 
ATDA \cite{liu2022atda} & 75.4 UA 76.2 WA \\ 
Area Attention \cite{xu2021speech} & 77.5 UA 79.3 WA \\ 
\hline
Our method & \textbf{80.3 UA 80.5 WA} \\
\hline
\end{tabular}
\end{center}
\end{table}

Next, we compare with other proposed models’ performance that used attention methods. 
For this, we chose our best results models that contained ECA blocks and STFT data augmentation. 
Table \ref{tab4} lists the results of the UA and WA evaluations.
All models compared with our method use spectrogram data and attention methods used for feature aggregation or extraction in the independent axis of the data.

As listed in Table \ref{tab4}, the proposed model shows a significantly better performance than the other models.
this is because the combination of deep CNN layers and ECA is an efficient structure for extracting emotional context.
In addition, the insufficient expressions of emotional features can be reinforced using the STFT augmentation method.
Therefore, for the SER, it is important to increase the representation of the emotional feature data and effectively learn the context within it.

\begin{figure}[!t]
\subfloat[\label{}]{
    \includegraphics[width=0.23\textwidth]{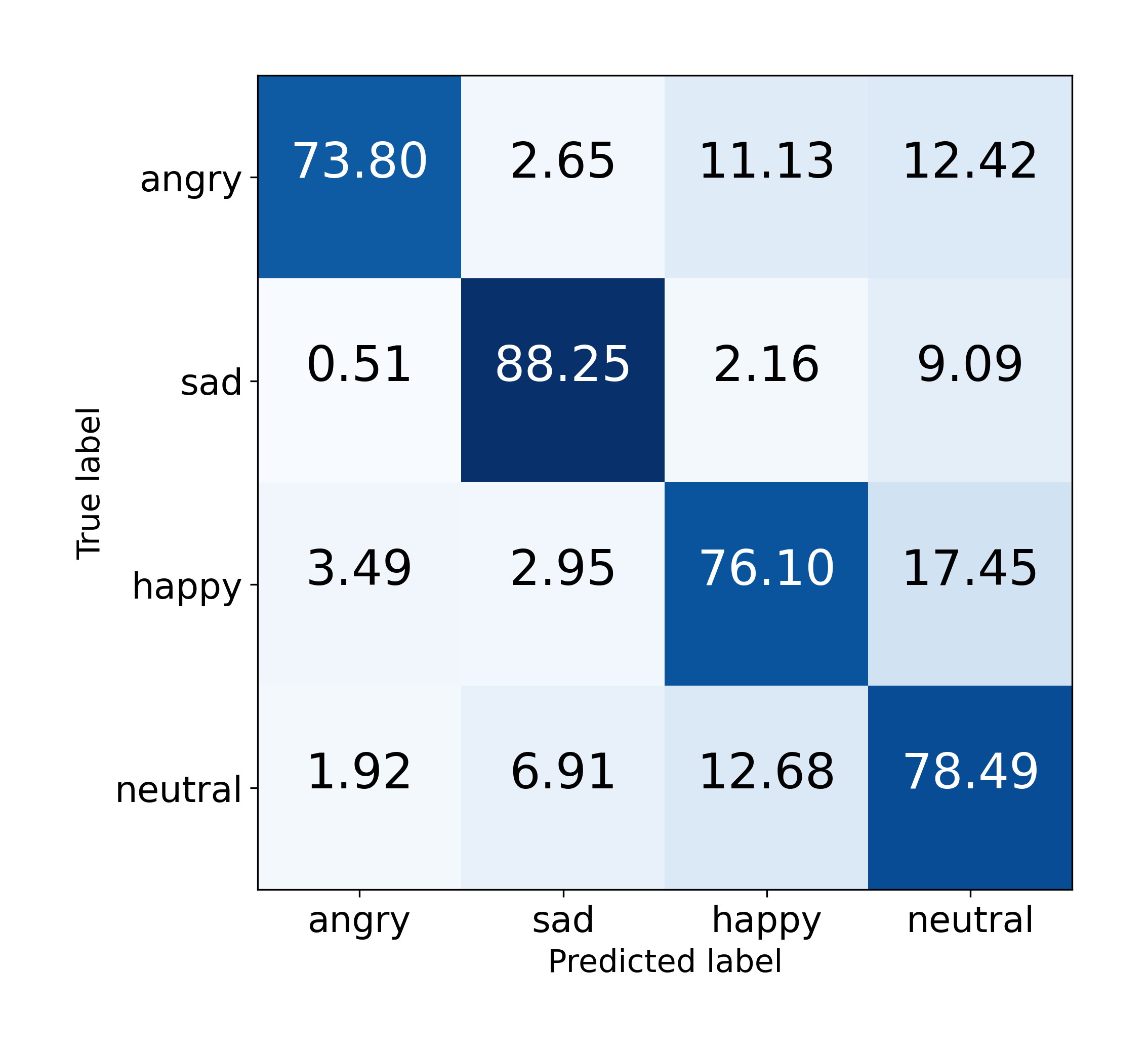}
}
\hfill
\subfloat[\label{}]{
    \includegraphics[width=0.23\textwidth]{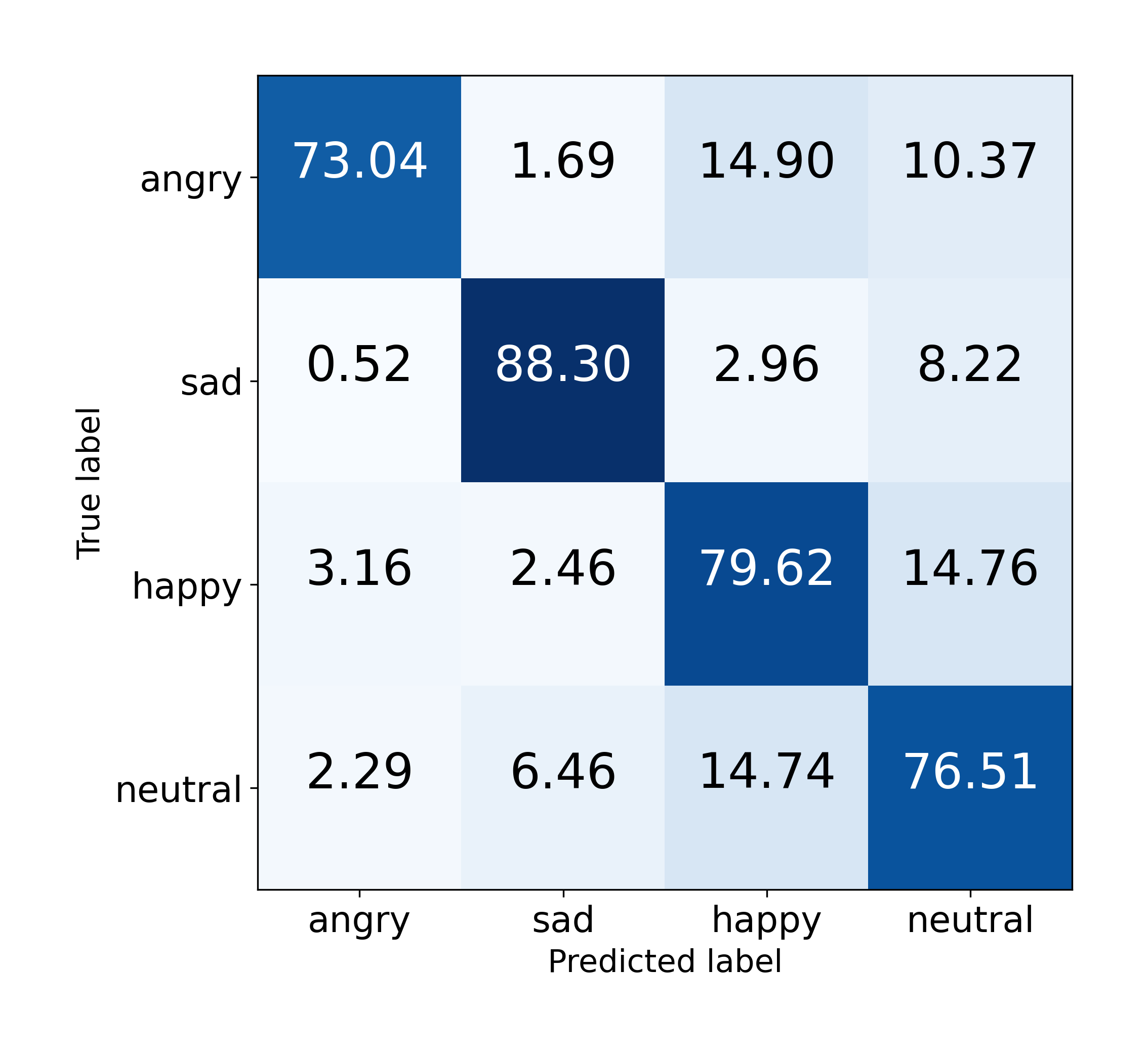}
}
\caption{The confusion matrices whether the ECA block is used or not in CNN-based model. (a) CNN-based model ($n=4$). (b) ECA block used in layer 5 and 6.}
\label{eca_use_confusion}
\end{figure}

\begin{figure}[!t]
\subfloat[\label{}]{
    \includegraphics[width=0.23\textwidth]{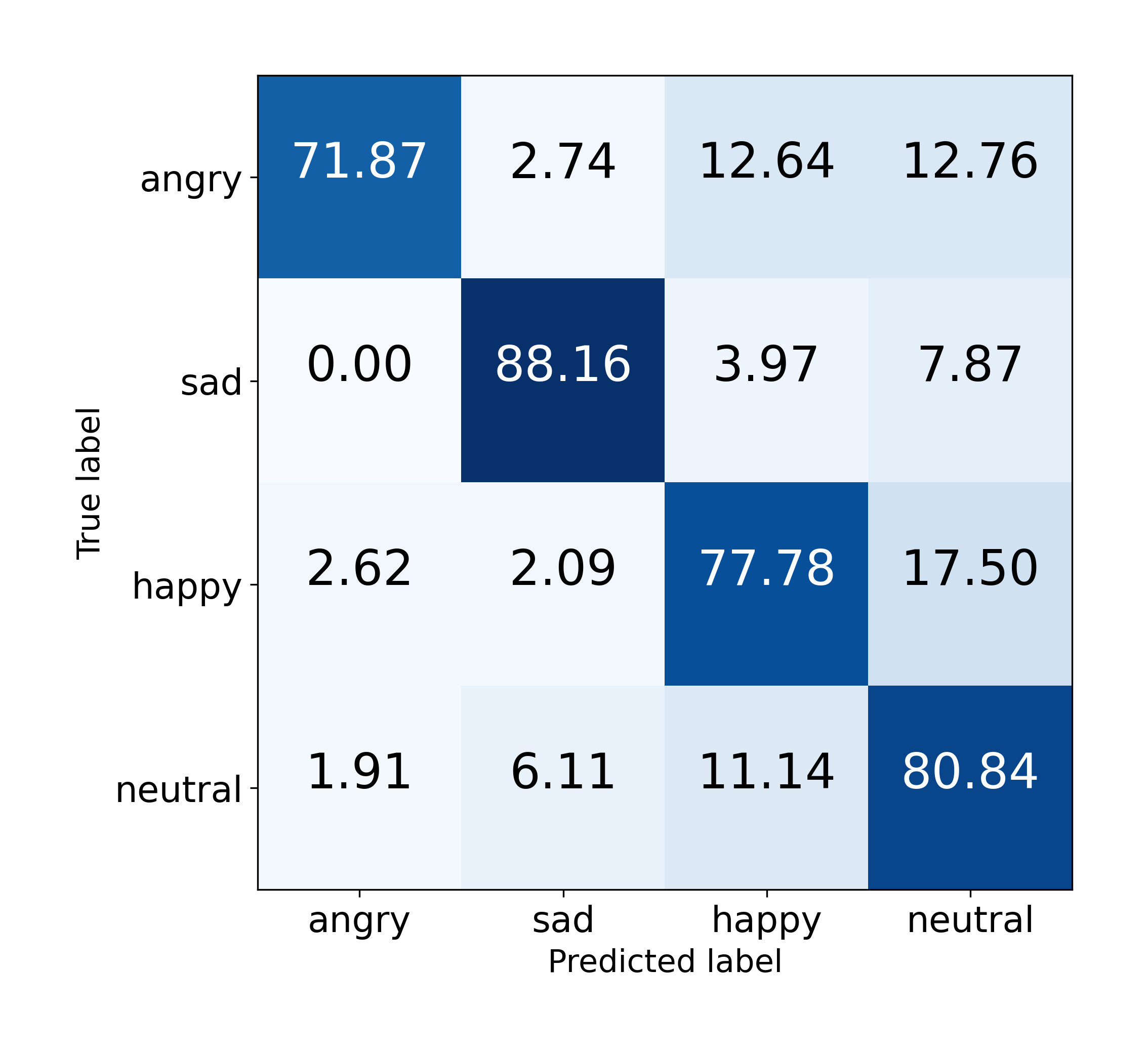}
}
\hfill
\subfloat[\label{}]{
    \includegraphics[width=0.23\textwidth]{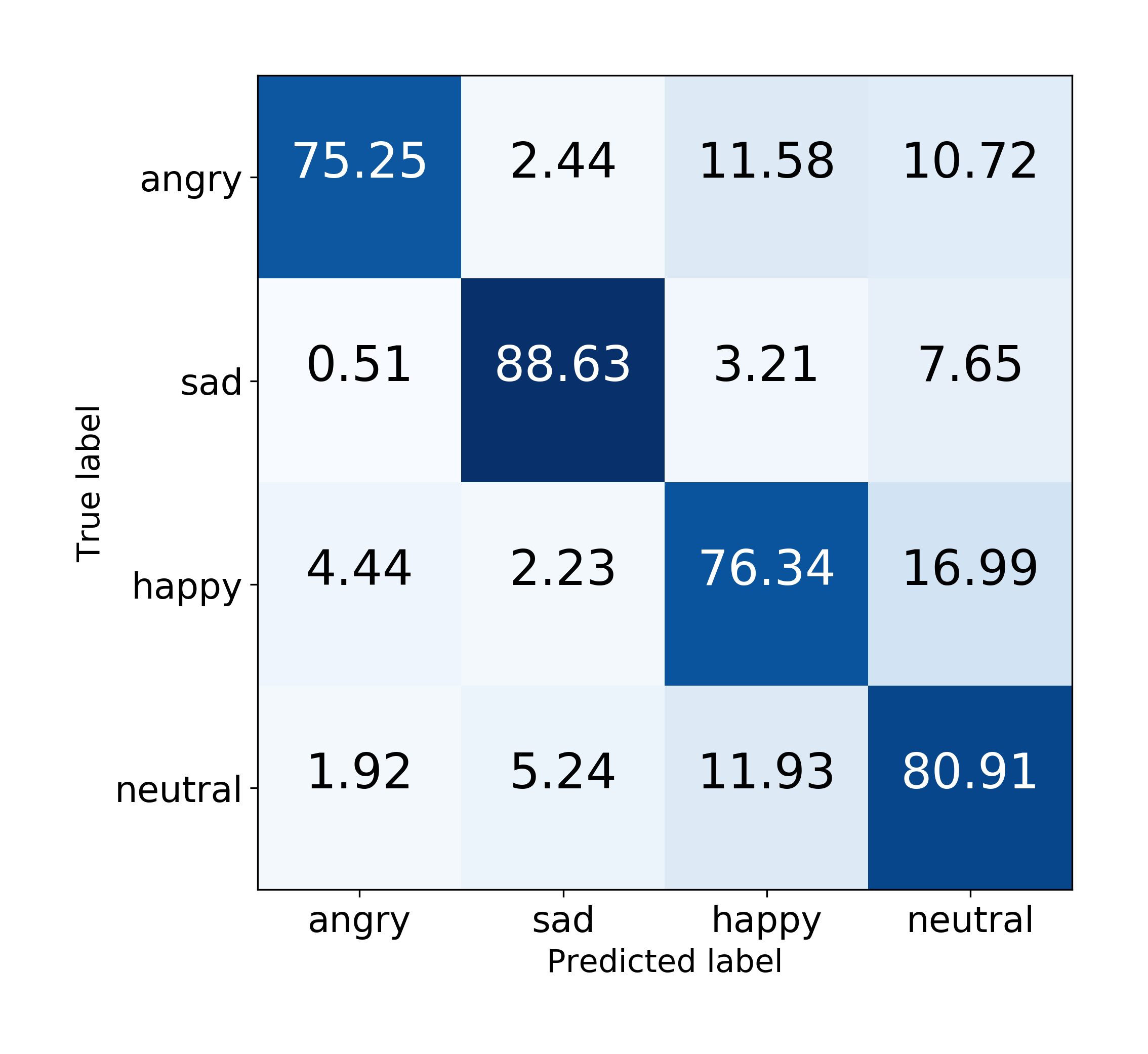}
}
\caption{The confusion matrices whether the ECA block is used or not in CNN-based model with STFT augmentation method. (a) CNN-based model ($n=4$). (b) ECA block used in layer 5 and 6.}
\label{aug_eca_use_confusion}
\end{figure}

\subsection{Analysis of the Ablation Studies}

\begin{table}
\begin{center}
\caption{ablation results}
\label{tab5}
\begin{tabular}{ l  l }
\hline
Method & Evaluation results \\
\hline
CNN-based model & 79.16 UA 79.27 WA 79.22 ACC \\ 
\hline
+ Original ECA block & 78.71 UA 78.80 WA 78.76 ACC \\ 
+ Our ECA block & \textbf{79.37 UA 79.68 WA 79.53 ACC} \\ 
\hline
+ STFT augmentation & 79.66 UA 80.50 WA 80.08 ACC \\ 
+ + Our ECA block & \textbf{80.28 UA 80.46 WA 80.37 ACC} \\ 
\hline
\end{tabular}
\end{center}
\end{table}

For a detailed analysis of the proposed methods, the results of the ablation models were compared.
Table \ref{tab5} lists performance depending on whether or not the ECA block and STFT data augmentation were used. 
From the overall results, we can observe that the emotion classification performance is improved when applying our proposed ECA block.
In particular, the effect of the ECA block can significantly improve performance when used together with the STFT data augmentation method.

Subsequently, we compared the classification performance for each emotion according to the ablation models. 
Fig. \ref{eca_use_confusion} and Fig. \ref{aug_eca_use_confusion} show the confusion matrices of the ablation models.
As shown in Fig. \ref{eca_use_confusion} and Fig. \ref{aug_eca_use_confusion}, you can see that in most models, the classification performance of angry and sadness was high; however, the classification performance for angry, happiness, and neutral tended to be low.
Compared with Fig. \ref{eca_use_confusion}(a), Fig. \ref{aug_eca_use_confusion}(b) shows that the classification performance of all emotions improved when the ECA and STFT augmentation methods were used.

Specifically, in Fig. \ref{eca_use_confusion}(a) and (b), the happiness classification accuracy improved by 3.5\%, and the neutral classification accuracy decreased by 2\% using the ECA.
It means that the ambiguity of classifying happiness and neutral was resolved through the ECA block.
As shown in Fig. \ref{aug_eca_use_confusion}, the neutral classification accuracy improved by approximately 2\% $\sim$ 4\%.
In particular, as shown in Fig. \ref{aug_eca_use_confusion}(b), both angry and neutral classification accuracies were significantly improved.

\begin{figure}[!t]
\subfloat[\label{}]{
    \includegraphics[width=0.5\textwidth]{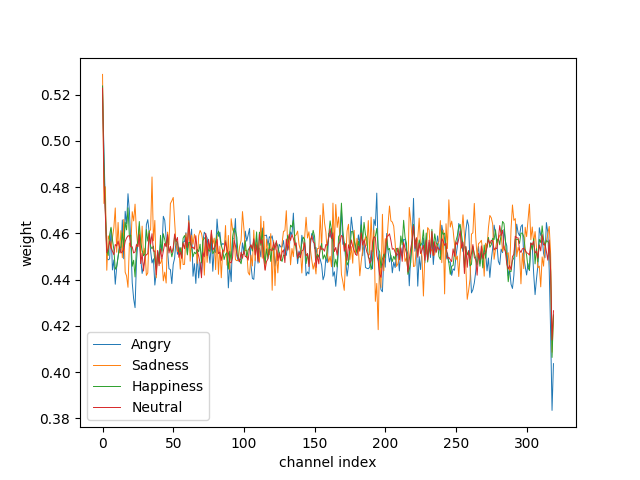}
}
\hfill
\subfloat[\label{}]{
    \includegraphics[width=0.5\textwidth]{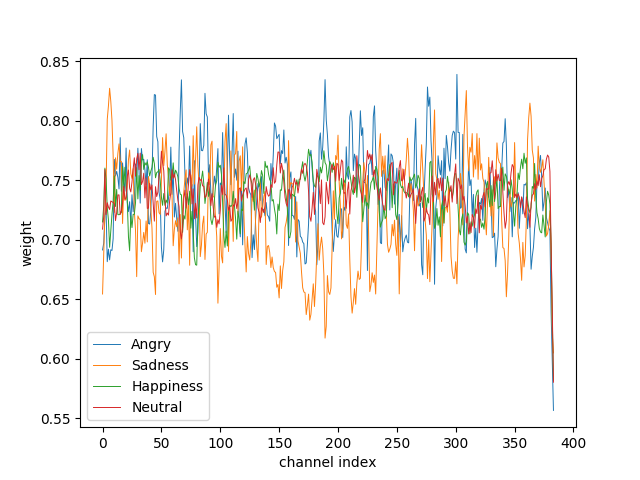}
}
\caption{The ECA's channel weight plots of each emotion class. (a) The channel weights in layer 5. (b) The channel weights in layer 6.}
\label{eca_channel_weight}
\end{figure}

The results show that channel feature extraction with the ECA blocks is effective for SER.
Subsequently, to understand how the ECA block works for classifying each emotion, we checked the channel weights for each emotion.
Fig. \ref{eca_channel_weight} shows a plot of the channel weight of the ECA blocks learned in the 5th and 6th layers when using the STFT augmentation method.
To plot the channel weights, the weights from the test set were averaged.

As shown in Fig. \ref{eca_channel_weight}(a), the ECA weights of the fifth layer are not related to any emotion.
However, as shown in Fig. \ref{eca_channel_weight}(b), the ECA weights of the last layer are noticeably different.
In particular, channel weights were distinguishable between angry (blue line) and sadness (orange line).
In addition, neutral (red line) is distinct from angry (blue line) and sadness (orange line).
However, it is slightly different from the neutral (red line) to happiness (green line), because it is difficult to distinguish between them.

In summary, it is difficult to distinguish between all emotions, especially angry, happiness, and neutral emotions. 
However, the proposed method can increase the accuracy of all emotions. 
Therefore, adopting channel attention in the CNN-based model can effectively extract the emotional context.

\section{Conclusion}
\label{conclusion}
This study proposes a promising and more effective preprocessing method and an ECA module for SER, offering the potential for significant advancements in the field.
Our experiment, conducted with eight different preprocessing datasets from the IEMOCAP corpus, revealed a significant finding: a spectrogram with a higher frequency resolution is more effective in training emotional features, providing valuable insight for future research in the field.
Our study is the first study to apply an ECA to the SER.
We achieved significantly better results than previous models by applying ECA to our CNN-based model with an effective preprocessing method.

Considering these results, correctly understanding the relationship between the channel features in the CNN structure can be a clue to understanding how to distinguish emotions.
However, the ECA is limited in that it only considers the relationship between neighboring channels.
In future work, we will look for attention structures that are efficient but can learn the relationships between channel features more broadly.
In addition, it is necessary to determine a better preprocessing method for emotion recognition by analyzing the frequency features associated with each emotion.


\newpage

\vspace{11pt}

\vfill

\end{document}